\documentstyle[12pt,epsfig,psfig]{article}
\begin{document}
\begin{center}
\large{\bf {Nuclear Physics of Double Beta Decay}}

\vspace {0.5cm}
\large{Petr Vogel}\footnote{Invited talk at the Workshop on Beta-Decay,
Strasbourg, France, March 1999}

\vspace{0.5cm}
\large{Caltech, Physics Dept. 161-33, Pasadena, CA 91125, USA}
\end{center}
\begin{abstract}
Study of the neutrinoless $\beta\beta$ decay allows us to put a stringent
limit on the effective neutrino Majorana mass, a quantity of fundamental
importance.  To test our ability to evaluate the nuclear matrix elements
that govern the decay rate,  it is desirable to be able to describe the
allowed two-neutrino decay. It is argued that only low-lying virtual 
intermediate states are important for that process, and thus it appears
that the large-scale shell model evaluation, free from the various
difficulties of QRPA, is the preferred method. In the $0\nu$ decay, 
it is shown that there is a substantial cancellation between the
paired and broken pair pieces of the nuclear wave function. Thus,
despite the popular claim to the contrary, one expects that the $0\nu$
rate is also sensitive to the details of nuclear structure. 
This is reflected in
the spread of the theoretical values, leading to the uncertainty
of about factor 2-3 in the deduced $\langle m_{\nu} \rangle$ limit.
Finally, brief comment about the $0\nu$ decay with the exchange
of a very heavy neutrino are made.
\end{abstract}

\section{Introduction}

Double beta decay has been long recognized as a powerful tool 
for the study of lepton conservation. The theoretical description 
of $\beta\beta$ decay involves particle physics
and nuclear structure physics. The latter is the topic 
of the present work. After a brief introduction and a review 
of the experimental situation I will describe
three distinct set of problems:

\begin{itemize}
\item $2\nu$ decay: the physics of the Gamow-Teller amplitudes.
\item $0\nu$ decay - exchange of light massive Majorana neutrinos:
no selection rules on multipoles, role of nucleon correlations, sensitivity
to nuclear models.
\item $0\nu$ decay - exchange of heavy neutrinos: physics 
of the nucleon-nucleon states at short distances.
\end{itemize}

Since the lifetimes of $\beta\beta$ decay are so long, the experimental
search has spawned a whole field of experiments requiring very low background.
During the last decade there has been enormous progress in the experimental 
study of the double beta decay. The $2\nu$ decay, with lifetimes of 
$10^{19} - 10^{21}$ y, has been now observed in ten cases, often repeatedly
and by different groups \cite{Morales}. 
The halflives are typically
determined to better than 10\% accuracy. Thus, 
the $2\nu$ decay has become a tool,
rather than an exotic curiosity.

The main experimental effort is concentrated, naturally, 
on the search for the $0\nu$ decay.  There, multikilogram sources 
with very low background are the state of the art at
present. Various techniques are being pursued: 
two large experimental efforts (Heidelberg-Moscow collaboration 
in Gran Sasso and the IGEX collaboration in Canfranc) 
employ enriched $^{76}$Ge in sets of detectors 
operated deep underground. The latest result \cite{Baudis}
is based on 41.55 kg y of exposure and  
halflife limit $T_{1/2} > 1.3 \times 10^{25}$y (90\% CL). 
Alternatively, the analysis of the 24.16 kg y of data with 
pulse shape measurement, and hence improved background
suppression, leads to even better limit of 
$T_{1/2} > 1.6 \times 10^{25}$y (90\% CL).
Note, that in the latter case the actual number of observed events 
is smaller than the expected background. When taking advantage of this, 
the authors of \cite{Baudis} report yet much
improved limit of $T_{1/2} > 5.7 \times 10^{25}$y (90\% CL). 
The latest IGEX limit is 
$T_{1/2} \ge 0.8 \times 10^{25}$y (90\% CL)\cite{IGEX}.
I will discuss the interpretation of this result in terms 
of the neutrino Majorana mass below in Section 3.

Other techniques which allow expansions to multikilogram
sources involve gas TPC ($^{136}$Xe in the Gotthardt tunnel \cite{Xe}),
electron tracking detectors combined with calorimeters
(NEMO \cite{Nemo}, ELEGANTS \cite{Ejiri}), and  
cryogenic bolometers \cite{Fiorini}.
These experiments will make it possible
(or currently already achieve) to reach neutrino
mass limit well below 1 eV.

In addition, there are plans for improvements of the limit 
to the 0.1 eV range by scaling up the source mass 
to about one ton quantities. Such experiments,
with either enriched $^{76}$Ge (GENIUS) or with large amount 
of cooled TeO$_2$ (CUORE) will require very large capital 
expenditures and running times of about
five years. The field of $\beta\beta$ decay then becomes competitive, in
complexity, manpower, and price, to large accelerator particle 
physics experiments. 

The Majorana neutrino mass $\langle m_{\nu} \rangle$ below 1 eV
is an important landmark. This has been stressed often,
e.g., in the work of Georgi and Glashow \cite{GG}. There, the authors
quote as ``established facts'' the discovery by the SuperKamiokande
collaboration of the $\nu_{\mu} \rightarrow \nu_{\tau}$ neutrino oscillations 
with $\Delta m^2 \simeq 10^{-3}$ eV$^2$ and  mixing angle 
$\sin^2 2 \theta \simeq 1$, and the solar neutrino deficit,
confirmed in five experiments, which involves $\nu_e \rightarrow \nu_x$
oscillations with $\Delta m^2 \le 10^{-5}$ eV$^2$. The
``tentative facts'' are based on theoretical prejudices and need
experimental confirmation. They are the existence of precisely three
massive Majorana neutrinos, and the assumption that these neutrinos
are responsible for the hot dark matter, leading to the conclusion that
$m_1 + m_2 + m_3 \sim 6  $ eV. Taking all of that together one is forced
to the wholly unexpected and somewhat bizarre conclusion that the three
neutrino flavors are essentially degenerate with masses of 
2 eV each (albeit with a sizable uncertainty in this value). 

It is now clear that if the study of the $0\nu$ $\beta\beta$ decay can without
doubt establish that  $\langle m_{\nu} \rangle < 1$ eV, this finding has
a profound consequences for the structure of the neutrino mixing matrix.
In particular, as Georgi and Glashow \cite{GG} argue, it would lead to maximum
mixing involving electron neutrinos also.

\section{Two neutrino decay}

This decay, characterized by the transformation of two neutrons into
two protons with the emission of two electrons and two $\bar{\nu}_e$,
does not violate any selection rules. Since the energies involved are
modest, the allowed approximation should be applicable, and the rate
is governed by the double Gamow-Teller matrix element 
\begin{equation}
M_{GT}^{2\nu} = \sum_m \frac{\langle f || \sigma \tau_+ || m \rangle
\times  \langle m || \sigma \tau_+ || i \rangle } { E_m - (M_i + M_f)/2 } ~'
\end{equation}
where $i, f$ are the ground states in the initial and final nuclei, 
and $m$ are the intermediate $1^+$ (virtual) states in the odd-odd nucleus.
The first factor in the numerator above represents the $\beta^+$
(or ($n,p$)) amplitude for the final nucleus, while the second one
represents the $\beta^-$ (or ($p,n$)) amplitude for the initial nucleus.
Thus, in order to correctly evaluate the $2\nu$ decay rate, we have to
know, at least in principle, {\it all} GT amplitudes for 
both $\beta^-$ and $\beta^+$ processes,
including their signs. The difficulty is that the $2\nu$ matrix element
exhausts a very small fraction ($10^{-5} - 10^{-7}$) of the double GT
sum rule \cite{double}, and hence it is sensitive to details of 
nuclear structure.

\begin{figure}[h]
\begin{center}
\mbox{\psfig{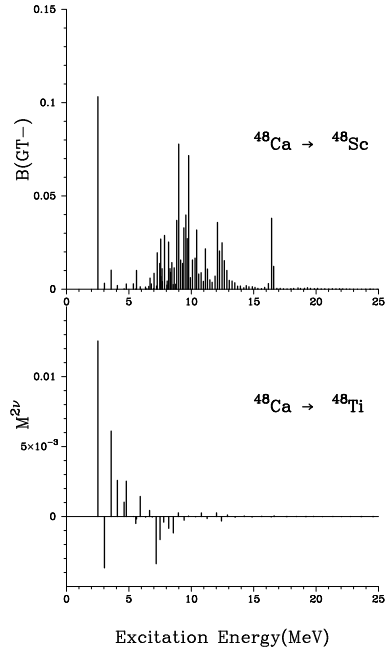}}
\caption{The $\beta^-$ strength (upper panel), and the
contributions to the $2\nu$ matrix element, eq. (1) (lower panel).
Both as the function of the excitation energy in $^{48}$Sc.}
\label{fig:ca48}
\end{center}
\end{figure}

Various approaches used in the evaluation of the $2\nu$ decay rate
have been reviewed recently by Suhonen and Civitarese \cite{SC}.
The Quasiparticle Random Phase Approximation (QRPA) has been the most
popular theoretical tool in the recent past. Its main ingredients, the
repulsive particle-hole spin-isospin interaction, and the attractive
particle-particle interaction, clearly play a decisive role in the
concentration of the $\beta^-$ strength in the giant GT resonance,
and the relative suppression of the $\beta^+$ strength and its 
concentration at low excitation energies. Together, these two 
ingredients are able to explain the suppression of the $2\nu$
matrix element when expressed in terms of the corresponding sum rule.

Yet, the QRPA is often criticized. Two ``undesirable'', and to some 
extent unrelated, features are usually quoted.
One is the extreme sensitivity of the decay rate to the
strength of the particle-particle force (often denoted as $g_{pp}$).
This decreases the predictive power of the method. The other one is the
fact that for a realistic value of  $g_{pp}$ the QRPA solutions are close
to their critical value (so-called collapse). This indicates a phase
transition, i.e., a rearrangement of the nuclear ground state. QRPA
is meant to describe small deviations from the unperturbed ground state,
and thus is not fully applicable near the point of collapse. 
Numerous approaches have been made to extend the range of validity of QRPA, 
see, e.g., Ref. \cite{SC}. The description
of all these generalizations is beyond the scope of this talk.  

\begin{table}[h]
\caption{ Experimental halflives of the $2\nu$ decay,
and the ratios of calculated to experimental halflives
(see text).}
\begin{center}
\begin{tabular}{lllc}  \hline\hline
   & $\frac{T_{1/2}^{calc}}{T_{1/2}^{exp}}$ 
& $\frac{T_{1/2}^{calc}}{T_{1/2}^{exp}}$
& $T_{1/2}$ (y) \\ 
Nucleus & QRPA &shell model & exp. \\  \hline 
$^{48}$Ca& -- & 0.91 & 4.3$\times 10^{19}$ \\
$^{76}$Ge& 0.71 &
1.44 &1.8$\times 10^{21}$ \\
$^{82}$Se& 1.5 & 0.46
&8.0$\times 10^{19}$ \\
$^{100}$Mo& 0.6 & --
&1.0$\times 10^{19}$ \\
$^{128}$Te& 0.27 & 0.25
&2.0$\times 10^{24}$ \\
$^{130}$Te& 0.27
& 0.29 &8.0$\times 10^{20}$ \\
$^{136}$Xe& $<$ 1.5 & $<$ 3.7
&$>$5.6$\times 10^{20}$ \\  \hline
\label{tab:2nu}
\end{tabular}
\end{center}
\end{table}

At the same time, detailed calculations show that 
the sum over the excited states in Eq.(1) converges 
quite rapidly \cite{EEV}. In fact, a few low-lying states usually
exhaust the whole matrix element. Thus, it is not really necessary to describe
all GT amplitudes; it is enough to describe correctly the $\beta^+$ and
$\beta^-$ amplitudes of the low-lying states, and include everything else in
the overall renormalization (quenching) of the GT strength. 
The situation is illustrated
in Fig. \ref{fig:ca48} modified from Ref. \cite{ca48sm}.

Nuclear shell model methods are presently capable of handling much larger
configuration spaces than even a few years ago. Thus, for many nuclei the
evaluation of the $2\nu$ rates within the 0$\hbar \omega$ shell model
space is feasible. (Heavy nuclei with permanent deformation, like
$^{150}$Nd and $^{238}$U remain, however, beyond reach.)
Using the interacting shell model avoids, 
naturally, the above difficulties of QRPA.
At the same time, the shell model is capable to predict, with the same method
and the same residual interaction, a wealth of spectroscopic data,
allowing a much better test of the predictive power.

To judge the degree of understanding of the $2\nu$ decay I show in
Table \ref{tab:2nu} the comparison with experiment of the initial Caltech
QRPA calculation \cite{Engel} and the modern shell model
\cite{Caurier97}. One can see that both methods are able, at least
in these cases, explain the $2\nu$ decay rates reasonably well, 
even though in the case of Te both methods underestimate the halflife
by a factor of about four.

\section{Neutrinoless decay: light Majorana neutrino}

In the neutrinoless decay the two electrons are
the only leptons emitted, and consequently their sum
energy is just the sharp nuclear mass difference.
This feature makes the experimental recognition of the
$0\nu$ decay much easier; it also results in a more favorable
phase space factor.

If one assumes that the $0\nu$ decay is caused by the exchange
(virtual) of a light Majorana neutrino between the two nucleons,
then several new features arise: a) the exchanged neutrino
has a momentum $q \sim 1/r_{nn} \simeq 50 - 100$ MeV
($r_{nn}$ is the distance between the decaying nucleons).
Hence, the dependence on the energy in the intermediate state
is weak and the closure approximation is applicable.
Also, b) since $qR > 1$ ($R$ is the nuclear radius), the expansion
in multipoles is not convergent, unlike in the $2\nu$ decay.
In fact, all possible multipoles contribute by a comparable amount.
Finally, c) the neutrino propagator results in a neutrino potential
of a relatively long range.

Thus, in order to evaluate the rate of the $0\nu$ decay, we need to
evaluate only the matrix element connecting the ground states $0^+$ of
the initial and final nuclei. Again, we can use the QRPA or the shell model.
Both calculations show that the features enumerated above are indeed
present. In addition, the QRPA typically shows less extreme dependence
on the particle-particle coupling constant $g_{pp}$, since the contribution
of the $1^+$ multipole is relatively small. The calculations also
suggest that for quantitatively correct results one has
to treat the short range nucleon-nucleon repulsion carefully, 
despite the long range of the neutrino potential.

Does that mean that the calculated matrix elements are insensitive to
nuclear structure? An answer to that question has obviously great
importance, since unlike the $2\nu$ decay, we cannot directly test whether
the calculation is correct or not. 

For simplicity, let us assume that the $0\nu$ $\beta\beta$ decay is
mediated only by the exchange of a light Majorana neutrino. The
relevant nuclear matrix element is then the combination 
$M_{GT}^{0\nu} - M_F^{0\nu}$, where the GT and F operators
change two neutrons into two protons, and contain the corresponding
operator plus the neutrino potential. One can express these matrix elements 
either in terms of the proton particle - neutron hole multipoles (i.e.
the usual beta decay operators) or in the multipoles coupling
of the exchanged pair, $nn$ and $pp$. 

\begin{figure}[h]
\begin{center}
\mbox{\epsfig{figure=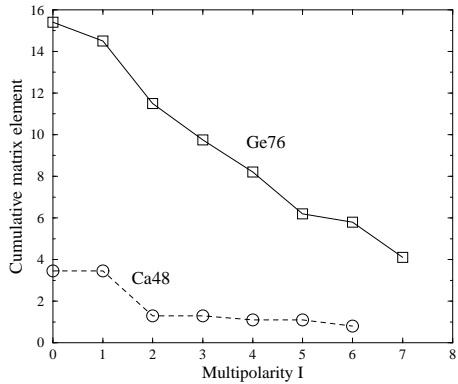,width=6cm}}
\caption{The cumulative contribution of the pair states
with the natural parity multipolarity to the $0\nu$ nuclear matrix 
element combination $M_{GT}^{0\nu} - M_F^{0\nu}$.
The full line is for $^{76}$Ge and the dashed line for $^{48}$Ca.}
\label{fig:0nu}
\end{center}
\end{figure}

When using the decomposition in the proton particle - neutron hole multipoles,
one finds that all possible multipoles (given the one-nucleon states near
the Fermi level) contribute, and the contributions have typically equal signs.
Hence, there does not seem to be much cancellation.
However, perhaps more physical is the decomposition into the exchanged
pair multipoles. There one finds, first of all, that only 
natural parity multipoles
($\pi = (-1)^I$) contribute noticeably. And there is a rather severe
cancellation. The biggest contribution comes from the $0^+$, i.e., the pairing
part. All other multipoles, related to higher seniority states, 
contribute with an opposite sign. The final matrix element is then 
a difference of the pairing and higher multipole 
(or broken pair $\equiv$ higher seniority) parts, and
is considerably smaller than either of them. This is illustrated in 
Fig. \ref{fig:0nu} where the cumulative effect is shown, i.e. the
quantity
\begin{equation}
M(I) = \sum_J^I \left[ M_{GT}^{0\nu}(J) - M_F^{0\nu}(J) \right]
\end{equation}
is displayed for $^{76}$Ge (from \cite{muto2}) and $^{48}$Ca
(from \cite{ca48sm}). Thus, the final result depends sensitively
on both the correct description of the pairing and on the admixtures
of higher seniority configurations in the corresponding initial
and final nuclei.

Since there is no objective way to judge which calculation is
correct, one often uses the spread between the calculated 
values as a measure of the theoretical uncertainty.
This is illustrated in Table \ref{tab:0nu}. There, I have chosen
two representative QRPA sets of results, the highly truncated
``classical'' shell model result of Haxton and Stephenson \cite{Haxton},
and the result of more recent shell  model calculation which is 
convergent for the set of single particle states chosen
(essentially 0$\hbar \omega$ space). 

For the most important case of $^{76}$Ge, the calculated rates
differ by a factor of 6-7. Since the effective neutrino mass
$\langle m_{\nu} \rangle$ is inversely proportional to the
square root of the lifetime, the experimental limit of
$1.6 \times 10^{25}$ y translates into limits of about 1 eV
according to \cite{Engel,Caurier97} and about 0.4 eV
according to \cite{Staudt,Haxton}. On the other hand, if one
would accept the more stringent limit of $5.7 \times 10^{25}$
\cite{Baudis}, even the more pessimistic matrix elements
restrict $\langle m_{\nu} \rangle < 0.5$ eV, hence the
scenario discussed by Georgi and Glashow \cite{GG}
is confirmed. Needles to say, a more objective measure of 
the theoretical uncertainty would be highly desirable. 

\begin{table}[h]
\caption{ Halflives in years calculated for $\langle m_{\nu} \rangle$
= 1 eV by various representative methods.}
\begin{center}
\begin{tabular}{r|rrll}  \hline\hline
   & QRPA \cite{Staudt}  
& QRPA \cite{Engel}
& SM \cite{Haxton} & SM \cite{Caurier97} \\  \hline
$^{76}$Ge& 2.3$\times10^{24}$ &
1.4$\times10^{25}$ &2.4$\times10^{24}$ & 1.7$\times10^{25}$ \\
$^{82}$Se& 6.0$\times10^{23}$ & 5.6$\times10^{24}$ &
8.4$\times10^{23}$ & 2.4$\times10^{24}$ \\
$^{100}$Mo & 1.3$\times 10^{24}$ & 1.9$\times 10^{24}$ & -- & -- \\
$^{130}$Te& 4.9$\times10^{23}$ & 6.6$\times10^{23}$ &
2.3$\times10^{23}$ & -- \\
$^{136}$Xe& 2.2$\times10^{24}$ & 3.3$\times10^{24}$ &
-- & 1.2$\times10^{25}$  \\  \hline
\end{tabular}
\end{center}
\label{tab:0nu}
\end{table}

\section{Neutrinoless decay: very heavy Majorana neutrino}

The neutrinoless $\beta\beta$ decay can be also mediated by the exchange
of a heavy neutrino. The decay rate is then inversely proportional 
to the square of the effective neutrino mass \cite{Vergados}. 
In this context it is particularly
interesting to consider the left-right symmetric model proposed by Mohapatra
\cite{Mohapatra}. In it, one can find a relation between 
the mass of the heavy neutrino $M_N$ and the mass of the right-handed 
vector boson $W_R$. Thus, the limit on the  $\beta\beta$ rate provides, 
within that specific model, a stringent
lower limit on the mass of $W_R$.

The process then involves the emission of the heavy $W_R^-$ by the first 
neutron, the vertex $W_R^- \rightarrow e^- + \nu_N$ followed
by $\nu_N  \rightarrow e^- + W_R^+$ with the absorption of the $W_R^+$
on the second neutron. Since all exchanged particles between the two neutrons 
are very heavy, the corresponding ``neutrino potential'' is of essentially
zero range. Hence, when calculating the nuclear matrix
element, one has to take into account carefully the short range 
nucleon-nucleon repulsion.

As long as we treat the nucleus as an ensemble of nucleons only, the only
way to have a nonvanishing nuclear matrix elements for the above process
is to treat the nucleons as finite size particles. 
In fact, that is the standard
way to approach the problem \cite{Vergados}; the nucleon size is described
by a dipole form factor with the cut-off parameter $\Lambda \simeq$  0.85 GeV.

\begin{figure}[h]
\begin{center}
\mbox{\psfig{figure=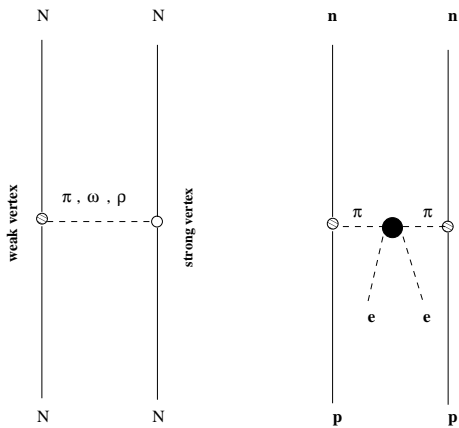,width=6cm}}
\caption{The Feynman graph description of the parity-violating
nucleon-nucleon force (left graph) and of the $\beta\beta$ decay with
the exchange of a heavy neutrino mediated by the pion exchange.}
\label{fig:moh}
\end{center}
\end{figure}

However, another way of treating the problem is possible, and already mentioned
in \cite{Vergados}. Let us recall how the analogous situation is treated
in the description of the parity-violating nucleon-nucleon force \cite{AH}.
There, instead of the weak (i.e., very short range) interaction of two nucleons,
one assumes that a meson ($\pi, \omega, \rho$) is emitted by one nucleon
and absorbed by another one. One of the vertices is the parity-violating one,
and the other one is the usual parity-conserving strong one. 
The corresponding range is then
just the meson exchange range, easily treated. The situation is schematically
depicted in the left-hand panel of Fig. \ref{fig:moh}. 
The analogy for $\beta\beta$
decay is shown in the right-hand graph. It involves two pions, and the
``elementary'' lepton number violating
$\beta\beta$ decay then involves a transformation of two pions
into two electrons. Again, the range is just the pion exchange range.
To my knowledge, no detailed evaluation of the corresponding graph was 
ever made (see, however, Ref.\cite{Savage}). 
It would be interesting to see if it would lead to a more or less
stringent limit on the mass of the $W_R$ than the treatment with form factors.

\vspace{1.5cm}

{\large{\bf Acknowledgment}}

It is my pleasant duty to thank Prof. Christianne Miehe 
for organizing the workshop and inviting me to Strasbourg. 
The permission of Etienne Caurier and Frederick Nowacki
to use their results is greatly appreciated. This work was supported
by the US Department of Energy under Grant No. DE-FG03-88ER-40397.

\end{document}